\documentclass[%
 reprint, 
superscriptaddress,
preprintnumbers,
nofootinbib,
 amsmath,amssymb,
 aps, physrev,
]{revtex4-2}

\usepackage{graphicx}
\usepackage{dcolumn}
\usepackage{bm}

\usepackage{relsize}
\usepackage{xcolor}

\usepackage{hyperref}
\hypersetup{
	colorlinks=true,
	linktoc=section,   
	citecolor=blue, 
	filecolor=blue,
	linkcolor=blue,
	urlcolor=blue,
	pdfborder={0 0 0.5},
	pdfmenubar=false,
	pdftoolbar=false
}

\begin{document}

\preprint{MPP--2025--218}

\title{\textbf{Instantons meet resonances: \\ Unifying two seemingly distinct approaches to quantum tunneling}}

\author{Björn~Garbrecht}%
\affiliation{Physik--Department T70, Technische Universität München, James--Franck--Stra{\ss}e 1, D--85748 Garching, Germany}

\author{Nils~Wagner}%
\email[Contact author: ]{nils.wagner@tum.de}
\affiliation{Physik--Department T70, Technische Universität München, James--Franck--Stra{\ss}e 1, D--85748 Garching, Germany}
\affiliation{Max--Planck--Institut für Physik (Werner--Heisenberg--Institut), Boltzmannstra{\ss}e 8, D--85748 Garching, Germany}

\date{\today}

\begin{abstract}
In the study of quantum-mechanical tunneling processes, numerous approaches have been developed to determine the decay rate of states initially confined within a metastable potential region. Virtually all analytical treatments, however, fall into one of two superficially unrelated conceptual frameworks: the resonant-state approach and the instanton method. Whereas the concept of resonant states and their associated decay widths is grounded in physical reasoning by capturing the regime of uniform probability decay, the instanton method lacks a comparably clear physical interpretation. We demonstrate the equivalence of the two approaches, revealing that the contour-deformation prescription in the functional integral put forward by Callan \& Coleman directly corresponds to the outgoing Gamow--Siegert boundary conditions defining resonant states.
\end{abstract}

\maketitle

\section{\label{sec:Introduction} Introduction}
\vspace{-0.2cm}

Originating from the seminal works of Langer~\cite{LangerCondensationPoint,LangerMetastability} and Callan \& Coleman~\cite{ColemanFateOfFalseVac1,CallanColemanFateOfFalseVac2}, the instanton method has become a standard tool for determining decay rates, especially in the infinite-dimensional context of quantum field theories~\cite{SherVacuumStabilitySM,Espinosa_VacSStability,IsidoriMetaStabilitySM,ButtazzoSMLifetime,SchwartzPrecisionDecayRate,SchwartzScaleInvInstantons}. Despite its remarkable success, the approach lacks a clear physical foundation that explains why it yields the sought-after result. In particular, the connection between this Euclidean formalism and the real-time dynamics underlying the interpretation of decay rates has so far remained conceptually unclear. Although recent works have made strides toward addressing this issue~\cite{SchwartzPrecisionDecayRate,TurokRealTimeTunneling,UnsalRealTimeInstantons,TanizakiLefschetz,LehnersComplexTimePaths,SchwartzDirectMethod,GarbrechtFunctionalMethods,MouRealTimeTunneling,HertzbergRealTimeTunneling,MatsuiRealTimePI,NishimuraRealTimeTunneling,BlumRealTimeTunneling,SteingasserFiniteTemp,SteingasserRealTimeInstantons,LawrenceRealTimeTunneling,FeldbruggeRealTimeTunneling,SteingasserExcitedStateSteadyonslong,SteingasserExcitedStateSteadyonsshort}, a clear first-principles understanding has so far been lacking~\cite{Wagner_CC_Resolution}.

In this letter, we outline how the instanton method relates to the notion of resonant states, which provide a physically grounded, idealized description of decaying wave functions~\cite{IntroductionGamovVectors,ResonancesIntroduction}. All subsequent developments are formulated for one-dimensional quantum-mechanical systems, for which the mathematical framework of resonant states attains its simplest form; extensions to higher dimensions and the field-theoretical case will be addressed in future work. 

This document is structured as follows: We begin by briefly reviewing the role of resonant states in decay-rate calculations, outlining their connection to the real-time dynamics governing quantum-mechanical probability decay. We then explain how these resonant states can be viewed as solutions to generalized eigenvalue problems, for which the boundary conditions are naturally specified in the complex plane rather than on the real axis. This less commonly emphasized perspective on resonant states proves particularly valuable in the present context. Focusing on the resonant eigenenergies that encode the desired decay rates, we subsequently outline how the spectrum of such generalized eigenvalue problems can be probed using functional methods, constituting the central result of our work. Finally, we leverage this newly developed toolset to obtain the ground-state decay rate $\Gamma_{\!\!\:0}$, thereby clarifying how this quantity arises within the traditional instanton method and establishing the sought-after connection between the previously distinct approaches. 

Excessively technical details and derivations are omitted in the present letter and are instead presented in full in the comprehensive companion paper~\cite{Wagner_CC_Resolution}.

\vspace{-0.15cm}

\section{Aspects of resonant states}
\vspace{-0.2cm}

We consider a potential $V(z)$ of the form shown in FIG. 1, featuring a metastable false vacuum (FV) region that permits quantum tunneling toward the true vacuum (TV). Preparing an initial wave function $\Psi_{T=0}(z)$ such that it is fully supported inside the FV region, one can study the time-dependence of the survival probability $\smash{P_{\scalebox{0.7}{\text{FV}}}(T)=\displaystyle{\scalebox{1.3}{$\int$}}_{\!\!\mathrm{FV}}\big\lvert \Psi(z,\!\!\:T) \big\rvert^2 \mathrm{d}z}$ under the unitary time evolution generated by the Hamiltonian $\widehat{H}=\hat{p}^2\!\!\:/(2m)+V(\hat{z})$. 

\begin{figure}[h]
\centering
\includegraphics[width=0.4\textwidth]{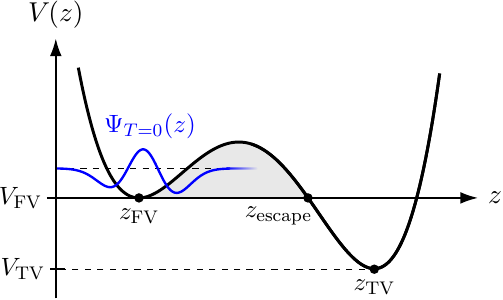}
\caption{Prototypical potential $V(z)$ for which one studies the survival probability of an initial state $\Psi_{T=0}(z)$, supported inside the FV region at the onset of the decay.}
\textbf{\label{fig:MetastablePotential}}
\end{figure}

In generic situations, three distinct temporal regimes emerge~\cite{KhalfinLateTimeBehavior,FondaDecayTheory,PeresDecayLaw}: At the onset of the decay, the initial wave function undergoes transient oscillations within the FV region before settling into a quasi-stationary configuration resembling an approximate eigenstate of a hypothetically stabilized FV basin. Once an approximately steady state has been reached, the long-lived intermediate temporal regime is characterized by uniform probability loss, giving rise to the familiar exponential decay dynamics. At sufficiently late times, the exponential decay law breaks down as back-reaction effects become non-negligible. Owing to its longevity, the intermediate temporal regime is of primary importance in most scenarios. Moreover, the uniformity of the decay during this period makes it particularly amenable to deeper theoretical analysis, since the inherently time-dependent problem can be effectively treated by time-independent methods.

Tracing back to the pioneering works of Gamow and Siegert~\cite{GamowAlphaDecay,SiegertRadiativeStates}, the long-lived quasi-stationary regime of exponential decay can be modeled by solving the time-\underline{in}dependent Schrödinger equation $\smash{\widehat{H}\Psi=E\Psi}$ with outgoing-wave boundary conditions, subsequently referred to as \emph{Gamow--Siegert boundary conditions}. Imposing the associated eigenfunctions of $\smash{\widehat{H}}$, dubbed \emph{resonant states}, to carry flux purely directed toward the TV region explicitly breaks the Hermiticity of $\smash{\widehat{H}}$, resulting in complex eigenvalues~\cite{IntroductionGamovVectors,ResonancesIntroduction}. With the imaginary parts of the eigenenergies $\smash{E_n^{\scalebox{0.7}{(\text{resonant})}}}$ being related to the outgoing flux, one finds the desired (ground-state) decay rate as 
\begin{equation}
    \Gamma_{\!\!\:0}=\frac{\text{outgoing flux}}{\text{FV population}}=-\:\!\frac{2}{\hbar}\,\mathrm{Im}\scalebox{1.2}{\big(}E_0^{\scalebox{0.7}{(\text{resonant})}}\scalebox{1.2}{\big)}\, .
\end{equation}

The question arises how these non-standard boundary conditions are to be instated effectively. In simplified WKB treatments, wave functions obtained by solving the Schrödinger equation in distinct regions of the potential are asymptotically matched in their respective overlap regions, with outgoing-wave boundary conditions being imposed beyond the classical escape point $z_{\scalebox{0.7}{\text{escape}}}$. However, studies of the analytic continuation of eigenvalue problems~\cite{BenderAnharmonicOscillator,BenderAnharmonicOscillator2,BenderAnalyticylContinuation} have provided a more suitable framework for imposing these modified boundary conditions by employing the notion of so-called \emph{Stokes wedges}, which we will briefly introduce in the subsequent section. 



\section{Generalized eigenvalue problems}

At this point, it is necessary to review some key aspects of Schrödinger-type differential equations. Studying the second-order ordinary differential equation (ODE) 
\begin{align}
	\widehat{H} \:\!\Psi(z)=\bigg\{\!-\frac{\hbar^2}{2m}\frac{\mathrm{d}^2}{\mathrm{d}z^2} + V(z)\bigg\} \,\Psi(z) = E\:\!\Psi(z)\, ,
	\label{eq:TimeIndependentSchrödingerEquation}
\end{align}
for a complex polynomial potential $V(z)$ of degree $\deg(V)=n$ and a fixed complex constant $E$, one characterizes the $n+2$ angular regions
\begin{align}
	\!\!S_k= \Bigg\{z\in\mathbb{C}\!\!\;:\!\!\;\scalebox{1.1}{\bigg\lvert}\, \mathrm{arg}(z)+\frac{\mathrm{arg}(c_n)-2\pi k}{n+2}\scalebox{1.1}{\bigg\lvert} \!\!\: <\frac{\pi}{n+2}\Bigg\} \, ,
    \label{eq:DefinitionStokesSectors}
\end{align}
with $c_n$ denoting the highest-order monomial coefficient of the polynomial $V(z)$, and $k$ being an integer in $\mathbb{Z}_{n+2}$. These \emph{Stokes wedges} turn out to be crucial for understanding the asymptotic structure of (non-constant) solutions $\Psi(z)$ to the ODE~\eqref{eq:TimeIndependentSchrödingerEquation}, as one can prove that, for $\lvert z\rvert\to\infty$, the magnitude $\big\lvert \Psi(z)\big\rvert$ either asymptotically decays or diverges within each of the $n+2$ individual sectors~\cite{SibuyaEigenvalueProblems}. In case the magnitude of $\Psi(z)$ approaches zero in an angular wedge $S_k$, this solution is said to be \emph{subdominant} in that particular sector. The crucial property we will exploit is that imposing subdominance of the solutions to the Schrödinger-type ODE~\eqref{eq:TimeIndependentSchrödingerEquation} in two non-adjacent Stokes wedges restricts the admissible energy values $E$ to a discrete, non-degenerate set~\cite{SibuyaEigenvalueProblems,FedoryukAsymptoticAnalysis,ShinPTEigenvalues,EremenkoODE}. Therefore, equation~\eqref{eq:TimeIndependentSchrödingerEquation} together with the aforementioned subdominance boundary conditions instated in two non-adjacent sectors grants a well-defined eigenvalue problem, possessing a discrete spectrum $\{E_\ell\}_{\scalebox{0.7}{$\ell\!\in\!\mathbb{N}$}}$.

\begin{figure}
\centering
\includegraphics[width=0.36\textwidth]{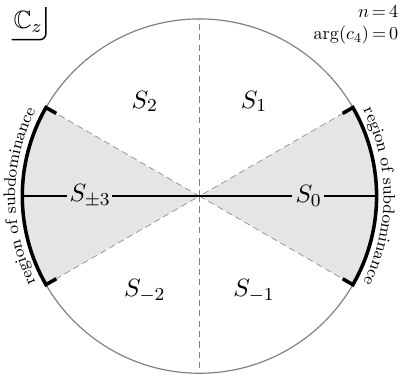}
\caption{Typical eigenvalue problem encountered in ordinary quantum mechanics, with the eigenfunctions defined in $L^2(\mathbb{R})$.}
\label{fig:TraditionalEVProblem}
\end{figure}
\begin{figure}
\centering
\includegraphics[width=0.35\textwidth]{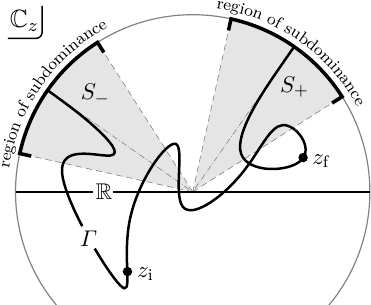}\hspace{0.225cm}
\caption{Generic boundary conditions imposed on the ODE~\eqref{eq:TimeIndependentSchrödingerEquation}, leading to a well-defined eigenvalue problem. Introducing an essentially arbitrary contour $\varGamma$ terminating in the desired regions of subdominance, the eigenvalue problem can be projected onto $\mathbb{R}$, allowing us to handle it with mostly standard quantum-mechanical tools.}
\label{fig:CustomContour_Gamma}
\end{figure}

Standard quantum mechanics offers the simplest example of this principle. For a real and globally stable potential, such as the one portrayed in FIG.~\ref{fig:MetastablePotential}, the real axis lies centered within two non-adjacent Stokes wedges, as illustrated in FIG.~\ref{fig:TraditionalEVProblem}. The usual requirement $\Psi_\ell(z)\in L^2(\mathbb{R})$ can therefore be equivalently satisfied by demanding that the eigenfunctions be subdominant in $S_0$ and $S_{\pm 3}$ in the example considered. However, the formalism at hand is more general, permitting the study of spectra arising from boundary data imposed in largely arbitrary sectors, e.g. $S_0$ and $S_2$. In due time, we will see how Gamow--Siegert boundary conditions fit into this framework. 

For now, let us stick to the most general case, exemplarily illustrated in FIG.~\ref{fig:CustomContour_Gamma}, for which the eigenvalue problem is defined in two arbitrary non-adjacent Stokes wedges $S_\pm$. It proves convenient to restrict the complex ODE~\eqref{eq:TimeIndependentSchrödingerEquation} underlying the eigenvalue problem to a suitably chosen one-dimensional complex contour $\varGamma$, which we demand to asymptotically terminate in the chosen sectors of subdominance $S_\pm$. Beware that this contour $\varGamma$, parametrized through the smooth surjection ${\gamma(\raisebox{1.1pt}{\scalebox{0.65}{$\bullet$}}):\mathbb{R}\to\varGamma}$, should not be confused with the desired decay rate $\Gamma_{\!\!\:0}$. Viewed on $\varGamma$, the eigenvalue problem attains the convenient form $\smash{\widehat{H}_\gamma \:\!\psi_\ell=E_\ell\:\! \psi_\ell}$, with the sought-after square-integrable eigenfunctions $\psi_\ell=\Psi_\ell\circ\gamma$ now defined in $L^2(\mathbb{R})$.\footnote{The concept of mapping non-Hermitian eigenvalue problems to the real line via a suitable complex contour is not new and has been previously discussed by Mostafazadeh, see e.g.~\cite{MostafazadehComplexContour,MostafazadehReview}.} At this point, all complexity has been hidden in the generally non-Hermitian, effective Hamiltonian $\smash{\widehat{H}_\gamma}$. Carefully accounting for the arising subtleties, the analysis of this modified eigenvalue problem proceeds using largely standard quantum-mechanical techniques. Defining an associated propagator which captures the probability amplitude to propagate from $z_\mathrm{i}$ to $z_\mathrm{f}$ along the contour $\varGamma$ in time $T$ and equating its spectral and path integral representations allows us to extract spectral information directly from a functional integral expression. A detailed derivation, utilizing techniques from non-Hermitian quantum mechanics~\cite{BariBiorthogonalBasis,SternheimBiorthogonalBasis,BrodyBiorthogonalBasis} and stochastic calculus~\cite{GrahamPathIntegralMethods1,GrahamPathIntegralMethods2,LangoucheFunctionalIntegration,dePireyPathIntegralMethods}, has been provided in the companion paper~\cite{Wagner_CC_Resolution}, arriving at the \underline{exact} relation
\begin{widetext}
\begin{equation}
	\mathlarger{\mathlarger{\int}}_{\scalebox{0.72}{$\mathcal{C}\big([0,T],\!\!\;\varGamma\big)$}}^{\substack{z(0)\:\!=\:\!z_\mathrm{i}\\[0.05cm] z(T)\:\!=\:\! z_\mathrm{f}\,}} \, \mathcal{D}[z]\,\exp\!\!\:\Bigg\{\!-\!\!\:\frac{1}{\hbar}\mathlarger{\int}_0^T\bigg[\frac{m\dot{z}(t)^2}{2}+V\!\!\;\scalebox{1.1}{\big(}z(t)\!\!\;\scalebox{1.1}{\big)}\bigg]\mathrm{d}t\Bigg\}\!\:=\mathlarger{\mathlarger{\sum}}_{\ell\!\:=\!\:0}^\infty \: \exp\!\bigg(\!\!-\!\!\:\frac{E_\ell T}{\hbar}\bigg) \:\! \Psi_\ell(z_\mathrm{i}) \, \Psi_\ell(z_\mathrm{f})\,\scalebox{1.1}{\bigg\{}\!\int_{\!\!\;\varGamma} \Psi_\ell(z)^2\,\mathrm{d}z\scalebox{1.1}{\bigg\}}^{\!\!\!\:-1} .
    \label{eq:Final_Relation_Propagator}
\end{equation}
\end{widetext}
As is customary, the functional integral is defined as the continuum limit of its associated discretized expression, comprising $N\to\infty$ individual contour integrals over $\varGamma$. By virtue of the Cauchy integral theorem, both the spectral and functional representations are invariant under deformations of $\varGamma$ that respect the regions of subdominance $S_\pm$ and the convergence of each discretized slice. Beware that the absence of complex conjugates in the otherwise formally conventional spectral representation is an essential feature.\footnote{For a real, confining Hermitian Hamiltonian $\smash{\widehat{H}}$, for which one can choose $\varGamma=\mathbb{R}$, all eigenfunctions may be taken real, so that conjugation is redundant; any residual phase is global and cancels upon proper normalization.} Meanwhile, the path integral only differs from the typically encountered quantum-mechanical propagator through its complexified functional integration contour $\mathcal{C}\big([0,T],\!\!\;\varGamma\big)$,\footnote{Similar generalizations had previously been conjectured for $\mathcal{PT}$-symmetric eigenvalue problems, see e.g.~\cite{Bender_PT_PathInt,Ai_PT_PathInt}.} with $\mathcal{C}(\mathcal{X},\mathcal{Y})$ denoting the function space of continuous maps from $\mathcal{X}$ to $\mathcal{Y}$. Thus, the chosen Stokes wedges $S_\pm$ defining the eigenvalue problem are ultimately encoded in the set of saddle points contributing to a semiclassical evaluation of the path integral~\eqref{eq:Final_Relation_Propagator} via the decomposition of its integration contour $\mathcal{C}\big([0,T],\!\!\;\varGamma\big)$ into \emph{Lefschetz thimbles}~\cite{PhamPicardLefschetz,HowlsPicardLefschetzTheory,WittenAnalyticContinuation,TanizakiLefschetz}. As we will see in the following section, relation~\eqref{eq:Final_Relation_Propagator} provides the foundation of the traditional instanton method, since the desired resonant states are selected by an appropriate choice of Stokes wedges $S_\pm$.

\begin{figure}[h]
\includegraphics[width=0.47\textwidth]{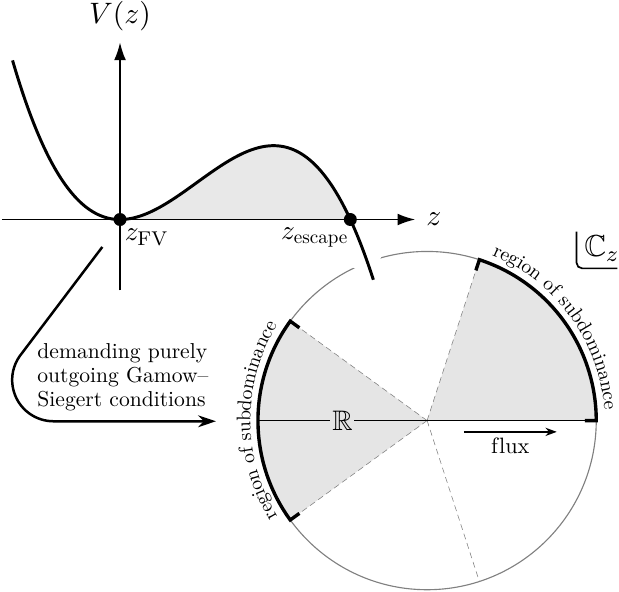}
\caption{Example of a real, unbounded cubic potential permitting Gamow--Siegert boundary conditions via an appropriate choice of Stokes wedges bordering the real axis in the spatial directions allowing for quantum tunneling.}
\label{fig:OddPoweredPolynomial}
\end{figure}

\begin{figure*}
\includegraphics[width=\textwidth]{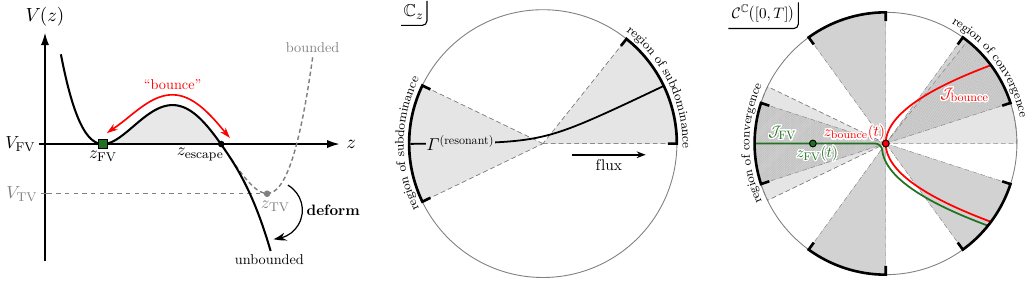}
\caption{\label{fig:Tunneling_Thimble_Structure_combined} (Left) Formal potential deformation required to naturally encode the desired Gamow--Siegert boundary conditions demanded to access resonant states. (Center) Emerging resonant-state eigenvalue problem defined in the modified, unbounded potential $\smash{V^{\scalebox{0.7}{(\text{deformed})}}(z)}$. (Right) Simplified thimble decomposition in the complexified function space $\mathcal{C}^\mathbb{C}\big([0,T]\big)$, schematically reducing the infinite-dimensional function space to a single field direction along which the FV and bounce trajectory are connected by the steepest-descent gradient flow. Accordingly, the saddle-point functions $z_{\scalebox{0.7}{\text{FV}}}(t)$ and $z_{\scalebox{0.7}{\text{bounce}}}(t)$ are represented as single points in the function space. Since the functional integration contour $\smash{\mathcal{C}\big([0,T],\varGamma^{\scalebox{0.7}{(\text{resonant})}}\:\!\big)}$ is deformed into the upper complex half-``plane'', the path integral~\eqref{eq:Ground_State_Energy_Resonant_FullySpelledOut} picks up half the bounce contribution, as previously indicated by Callan \& Coleman~\cite{CallanColemanFateOfFalseVac2}. As the system lies on a Stokes line, the residual contributions from the bounce thimble in the lower complex half-plane are canceled by its overlap with the FV thimble. To render all thimbles fully visible, this aforementioned overlap has been lifted.}
\end{figure*}

\section{Resonant states revisited}

Having established how to access the spectrum of generalized eigenvalue problems through a functional integral formulation, we can now return to examine how resonant states can be captured within the established framework. To this end, note that at the interface between a Stokes wedge where subdominance is imposed and its neighboring sectors, the solution becomes (asymptotically) purely oscillatory, reflecting the transition between decaying and growing behavior. Gamow--Siegert boundary conditions thus arise precisely when the real axis borders a sector in which subdominance is imposed~\cite{BenderAnharmonicOscillator,BenderAnharmonicOscillator2,BenderAnalyticylContinuation,MarinoAdvancedQM}. This situation occurs for real potentials that are unbounded in one (or both) spatial directions, as schematically illustrated in FIG.~\ref{fig:OddPoweredPolynomial}. Depending on whether subdominance is required in the sector above or below the real axis, one obtains purely incoming-wave or purely outgoing-wave boundary conditions, corresponding to resonant or anti-resonant states respectively.

To capture the desired resonances in a globally stable potential such as the one shown in FIG.~\ref{fig:MetastablePotential}, the potential must first be suitably deformed to realize the desired boundary conditions. Requiring a purely outgoing flux beyond the classical escape point $z_{\scalebox{0.7}{\text{escape}}}$, the potential has to be rendered unstable past $z_{\scalebox{0.7}{\text{escape}}}$, as shown in the left panel of FIG.~\ref{fig:Tunneling_Thimble_Structure_combined}.\footnote{Note that when utilizing semiclassical treatments to obtain the decay rate in quantum-mechanical settings, the decay rate is insensitive to the precise manner in which the potential deformation is introduced. This can be argued by the fact that the functional integral~\eqref{eq:Ground_State_Energy_Resonant_FullySpelledOut}, in the formal limit $\hbar\to 0^+$, only probes an infinitesimal neighborhood of the relevant saddle points.} In that way, the eigenvalue problem associated with the sought-after resonant states is defined in the deformed potential $\smash{V^{\scalebox{0.7}{(\text{deformed})}}(z)}$ as well as on a generalized contour $\smash{\varGamma^{\scalebox{0.7}{(\text{resonant})}}\neq \mathbb{R}}$, granting the desired Gamow--Siegert boundary conditions as shown in the central panel of FIG.~\ref{fig:Tunneling_Thimble_Structure_combined}. As is evident from equation~\eqref{eq:Final_Relation_Propagator}, these crucial features directly carry over to the path integral representation of the propagator, with the functional integration contour now given by $\mathcal{C}\big([0,T],\varGamma^{\scalebox{0.7}{(\text{resonant})}}\big)$. By conveniently choosing the otherwise arbitrary endpoints to be $z(0)=z(T)=z_{\scalebox{0.7}{\text{FV}}}$ to simplify the eventual semiclassical evaluation of the path integral, one can project out the desired resonant ground-state energy from the late-time behavior of the associated propagator~\eqref{eq:Final_Relation_Propagator}, yielding
\begin{widetext}
\begin{align}
    E_0^{\scalebox{0.7}{(\text{resonant})}}&=-\hbar \lim_{T\to\infty}\!\!\: \left(\!\!\;\frac{1}{T} \!\:\rule{0pt}{1.1cm}\log\!\!\!\;\left[\rule{0pt}{0.95cm}\smash{\mathlarger{\mathlarger{\int}}_{\scalebox{0.72}{$\mathcal{C}\big([0,T],\!\!\;\varGamma^{\scalebox{0.7}{(\text{resonant})}}\big)$}}^{\substack{z(0)\:\!=\:\!z_\mathrm{FV}\\[0.05cm] z(T)\:\!=\:\!z_\mathrm{FV}\,}}} \,\!\!\: \mathcal{D}[z]\, \exp\!\!\:\Bigg\{\!\!\!\:-\!\!\:\frac{1}{\hbar}\mathlarger{\int}_0^T\bigg[\frac{m\:\!\dot{z}(t)^2}{2}+V^{\scalebox{0.7}{(\text{deformed})}}\!\!\:\scalebox{1.1}{\big(}z(t)\!\!\;\scalebox{1.1}{\big)}\bigg]\mathrm{d}t\Bigg\}\right]\!\!\;\right) . \label{eq:Ground_State_Energy_Resonant_FullySpelledOut} 
\end{align}
\end{widetext}

Extracting the imaginary part of this FV-to-FV transition amplitude~\eqref{eq:Ground_State_Energy_Resonant_FullySpelledOut} via a semiclassical approximation is precisely the essence of the traditional instanton method.\footnote{To this end, note that the initial assumption of uniform decay corresponds to requiring a large barrier height and width, which ensures slow decay and allows the leading term in the formal $\hbar$-expansion of equation~\eqref{eq:Ground_State_Energy_Resonant_FullySpelledOut} to provide an excellent approximation to the desired ground-state decay rate $\Gamma_{\!\!\: 0}$.} Due to being defined within the unbounded potential $\smash{V^{\scalebox{0.7}{(\text{deformed})}}(z)}$, the functional integral~\eqref{eq:Ground_State_Energy_Resonant_FullySpelledOut} solely receives contributions from the trivial FV trajectory $z_{\scalebox{0.7}{\text{FV}}}(t)=z_{\scalebox{0.7}{\text{FV}}}$ as well as the traditional ``bounce'' motion $z_{\scalebox{0.7}{\text{bounce}}}(t)$, depicted in the left panel of FIG.~\ref{fig:Tunneling_Thimble_Structure_combined}. Additional saddle points probing the initially present true vacuum of the theory~\cite{PatrascioiuComplexTime,SchwartzPrecisionDecayRate}, nowadays often dubbed ``shot'' motions, are naturally absent. Moreover, due to $\varGamma^{\scalebox{0.7}{(\text{resonant})}}$ being deformed into the upper complex half-plane, the functional integration contour $\mathcal{C}\big([0,T],\varGamma^{\scalebox{0.7}{(\text{resonant})}}\big)$ follows suit, thereby explaining from first principles why only half of the bounce contribution enters the eventual result, as illustrated in the simplified steepest-descent thimble decomposition shown in the right panel of FIG.~\ref{fig:Tunneling_Thimble_Structure_combined}. To this end, note that the overlap region between the lightly shaded sectors of subdominance $S_\pm$, specifying the boundary conditions of the underlying eigenvalue problem, and the dark-gray sectors, illustrating the regions for which each discretized slice of the functional integral converges, precludes the real function space $\mathcal{C}\big([0,T],\mathbb{R}\big)$ from being an admissible integration contour. Although this contour-deformation was correctly recognized by Callan \& Coleman~\cite{CallanColemanFateOfFalseVac2}, its rigorous justification becomes clear only within the presented framework. 

Beware that the identical reasoning applies to variants of the instanton method for which the ground-state decay rate $\Gamma_{\!\!\: 0}$ is projected out using the (Euclidean) partition function $\smash{\mathrm{tr}\big(e^{-\widehat{H}_\gamma T/\hbar}\big)}$, which arises from equality~\eqref{eq:Final_Relation_Propagator} by choosing coincident endpoints and integrating them over $\varGamma^{\scalebox{0.7}{(\text{resonant})}}$. As before, the arising functional integral over periodic configurations ranges over the function space $\mathcal{C}\big([0,T],\varGamma^{\scalebox{0.7}{(\text{resonant})}}\big)$, with $\varGamma^{\scalebox{0.7}{(\text{resonant})}}$ chosen appropriately to instate the demanded Gamow--Siegert boundary conditions,\footnote{As noted after equation~\eqref{eq:Final_Relation_Propagator}, the arising result is independent of the specific choice of $\varGamma^{\scalebox{0.7}{(\text{resonant})}}$, so long as the contour terminates in the desired regions of subdominance and convergence.} thereby encoding the contour-prescription put forward by Callan \& Coleman~\cite{CallanColemanFateOfFalseVac2}. Additionally, relation~\eqref{eq:Final_Relation_Propagator} also provides a means to extract the decay rates associated to excited resonant states by weighting the propagator~\eqref{eq:Final_Relation_Propagator} with suitably chosen approximate eigenfunctions, as explicitly demonstrated in Ref.~\cite{WagnerExcitedStateTunneling}.

\section{Conclusion}
In this letter, we have elucidated the connection between the functional instanton approach and the concept of resonant states. By answering this long-standing question, we also resolve several conceptual ambiguities. To this end, one notes that resonant states, which precisely capture the relevant decay rates, arise from an eigenvalue problem formulated in the complex plane, for which a consistent path integral representation can be constructed in close analogy with ordinary quantum mechanics~\cite{Wagner_CC_Resolution}. The defining feature of the arising functional integral~\eqref{eq:Ground_State_Energy_Resonant_FullySpelledOut} is its complexified integration contour $\mathcal{C}\big([0,T],\varGamma^{\scalebox{0.7}{(\text{resonant})}}\big)$, reflecting the outgoing Gamow--Siegert boundary conditions characteristic of resonant states. Appropriately decomposed as a sum over Lefschetz thimbles, this generalized integration contour precisely implements the prescription originally proposed by Callan \& Coleman~\cite{CallanColemanFateOfFalseVac2}, ensuring that the path integral~\eqref{eq:Ground_State_Energy_Resonant_FullySpelledOut} receives imaginary contributions corresponding to half of the classical instanton motion. Within the provided framework, one naturally understands why certain shot-like trajectories do not contribute to false vacuum decay, why the integration contour must be taken in the upper rather than the lower complex half-plane, and why the resulting quantity computed by the traditional instanton method indeed corresponds to the ground-state decay rate $\Gamma_{\!\!\: 0}$. Moreover, our analysis clarifies how field configurations with negative modes can enter the result for the decay rate, despite their inability to contribute to the semiclassical expansion of a Euclidean path integral defined over a real function space.

The successful understanding of tunneling transitions via instantons in the given one-dimensional case strongly suggests that this standard functional treatment should equally apply to false vacuum decay in field theory. In that context, however, the lack of a readily applicable notion of resonant states renders a first-principles derivation an intriguing open problem.

\begin{acknowledgments}
N.W. thanks Soo-Jong Rey for preliminary discussions that sparked the ideas presented in this work. N.W. gratefully acknowledges support from the German Academic Scholarship Foundation, the Marianne--Plehn-Program of the Elite Network of Bavaria, and the International Max Planck Research School on Elementary Particle Physics (IMPRS EPP).
\end{acknowledgments}

\bibliography{LetterVersion}

@article{LangerCondensationPoint,
    author = "Langer, J. S.",
    title = "{Theory of the condensation point}",
    doi = "10.1016/0003-4916(67)90200-X",
    journal = "Annals Phys.",
    volume = "41",
    number = "1",
    pages = "108--157",
    year = "1967",
    issn = {0003-4916}
}

@article{IsidoriMetaStabilitySM,
    author = "Isidori, Gino and Ridolfi, Giovanni and Strumia, Alessandro",
    title = "{On the metastability of the Standard Model vacuum}",
    eprint = "hep-ph/0104016",
    archivePrefix = "arXiv",
    doi = "10.1016/S0550-3213(01)00302-9",
    journal = "Nucl. Phys. B",
    volume = "609",
    number = {3},
    pages = "387--409",
    year = "2001"
}

@article{ButtazzoSMLifetime,
    author = "Buttazzo, Dario and Degrassi, Giuseppe and Giardino, Pier Paolo and Giudice, Gian F. and Sala, Filippo and Salvio, Alberto and Strumia, Alessandro",
    title = "{Investigating the near-criticality of the Higgs boson}",
    eprint = "1307.3536",
    archivePrefix = "arXiv",
    primaryClass = "hep-ph",
    doi = "10.1007/JHEP12(2013)089",
    journal = "JHEP",
    volume = "2013",
    number = {12},
    pages = "089",
    year = "2013"
}

@article{SchwartzScaleInvInstantons,
    author = "Andreassen, Anders and Frost, William and Schwartz, Matthew D.",
    title = "{Scale-invariant instantons and the complete lifetime of the standard model}",
    eprint = "1707.08124",
    archivePrefix = "arXiv",
    primaryClass = "hep-ph",
    doi = "10.1103/PhysRevD.97.056006",
    journal = "Phys. Rev. D",
    volume = "97",
    number = "5",
    pages = "056006",
    year = "2018",
    month = {Mar},
    publisher = {American Physical Society}
}

@article{NishimuraRealTimeTunneling,
    author = "Nishimura, Jun and Sakai, Katsuta and Yosprakob, Atis",
    title = "{A new picture of quantum tunneling in the real-time path integral from Lefschetz thimble calculations}",
    eprint = "2307.11199",
    archivePrefix = "arXiv",
    primaryClass = "hep-th",
    doi = "10.1007/JHEP09(2023)110",
    journal = "JHEP",
    volume = "2023",
    number = "9",
    pages = "110",
    year = "2023",
    month = "7"
}

@article{LangerMetastability,
    author = "Langer, J. S.",
    title = "{Statistical theory of the decay of metastable states}",
    doi = "10.1016/0003-4916(69)90153-5",
    journal = "Annals Phys.",
    volume = "54",
    number = {2},
    pages = "258--275",
    year = "1969",
    issn = {0003-4916}
}

@article{ColemanFateOfFalseVac1,
    author = "Coleman, Sidney R.",
    title = "{The fate of the false vacuum. I. Semiclassical theory}",
    doi = "10.1103/PhysRevD.15.2929",
    journal = "Phys. Rev. D",
    volume = "15",
    number = {10},
    pages = "2929--2936",
    year = "1977",
    publisher = {American Physical Society},
    note = "[Erratum: \href{https://doi.org/10.1103/PhysRevD.16.1248}{\emph{Phys. Rev. D}, vol. \emph{16} \!\!\:(4), p. 1248 (1977)}]"
}

@article{CallanColemanFateOfFalseVac2,
    author = "Callan, Jr., Curtis G. and Coleman, Sidney R.",
    title = "{The fate of the false vacuum. II. First quantum corrections}",
    doi = "10.1103/PhysRevD.16.1762",
    journal = "Phys. Rev. D",
    volume = "16",
    number = {6},
    pages = "1762--1768",
    year = "1977",
    publisher = {American Physical Society}
}

@incollection{PhamPicardLefschetz,
    title = "{Vanishing homologies and the $n$ variable saddlepoint method}",
    author = {Frédéric Pham},
    series = {Proc. Symp. Pure Math.},
    publisher = {American Mathematical Society},
    volume = {40},
    number = {2},
    pages = {319--333},
    year = {1983},
    title = {Singularities}
}

@article{HowlsPicardLefschetzTheory,
    author = {E. Delabaere and C. J. Howls},
    title = {{Global asymptotics for multiple integrals with boundaries}},
    volume = {112},
    journal = {Duke Math. J.},
    doi = {10.1215/S0012-9074-02-11221-6},
    number = {2},
    publisher = {Duke University Press},
    pages = {199--264},
    year = {2002}
}

@article{WittenAnalyticContinuation,
    author = "Witten, Edward",
    editor = "Andersen, Joergen E. and Boden, Hans U. and Hahn, Atle and Himpel, Benjamin",
    title = "{Analytic Continuation Of Chern-Simons Theory}",
    eprint = "1001.2933",
    archivePrefix = "arXiv",
    primaryClass = "hep-th",
    journal = "AMS/IP Stud. Adv. Math.",
    volume = "50",
    pages = "347--446",
    year = "2011"
}

@article{TurokRealTimeTunneling,
    author = "Turok, Neil",
    title = "{On quantum tunneling in real time}",
    eprint = "1312.1772",
    archivePrefix = "arXiv",
    primaryClass = "quant-ph",
    doi = "10.1088/1367-2630/16/6/063006",
    journal = "New J. Phys.",
    volume = "16",
    number = {6},
    pages = "063006",
    year = "2014",
    month = {jun},
    publisher = {IOP Publishing},
}

@MISC{UnsalRealTimeInstantons,
    author = "Cherman, Aleksey and Unsal, Mithat",
    title = "{Real-Time Feynman Path Integral Realization of Instantons}",
    eprint = "1408.0012",
    archivePrefix = "arXiv",
    primaryClass = "hep-th",
    month = "7",
    year = "2014"
}

@article{TanizakiLefschetz,
    author = "Tanizaki, Yuya and Koike, Takayuki",
    title = "{Real-time Feynman path integral with Picard-Lefschetz theory and its applications to quantum tunneling}",
    eprint = "1406.2386",
    archivePrefix = "arXiv",
    primaryClass = "math-ph",
    doi = "10.1016/j.aop.2014.09.003",
    journal = "Annals Phys.",
    volume = "351",
    pages = "250--274",
    year = "2014",
    month = {dec},
}

@article{LehnersComplexTimePaths,
    author = "Bramberger, Sebastian F. and Lavrelashvili, George and Lehners, Jean-Luc",
    title = "{Quantum tunneling from paths in complex time}",
    eprint = "1605.02751",
    archivePrefix = "arXiv",
    primaryClass = "hep-th",
    doi = "10.1103/PhysRevD.94.064032",
    journal = "Phys. Rev. D",
    volume = "94",
    number = "6",
    pages = "064032",
    year = "2016",
    month = {Sep},
    publisher = {American Physical Society}
}

@article{SchwartzDirectMethod,
    author = "Andreassen, Anders and Farhi, David and Frost, William and Schwartz, Matthew D.",
    title = "{Direct Approach to Quantum Tunneling}",
    eprint = "1602.01102",
    archivePrefix = "arXiv",
    primaryClass = "hep-th",
    doi = "10.1103/PhysRevLett.117.231601",
    journal = "Phys. Rev. Lett.",
    volume = "117",
    number = "23",
    pages = "231601",
    year = "2016",
    month = {Nov}
}

@article{SchwartzPrecisionDecayRate,
    author = "Andreassen, Anders and Farhi, David and Frost, William and Schwartz, Matthew D.",
    title = "{Precision decay rate calculations in quantum field theory}",
    eprint = "1604.06090",
    archivePrefix = "arXiv",
    primaryClass = "hep-th",
    doi = "10.1103/PhysRevD.95.085011",
    journal = "Phys. Rev. D",
    volume = "95",
    number = "8",
    pages = "085011",
    year = "2017",
    month = {Apr},
    publisher = {American Physical Society}
}

@article{GarbrechtFunctionalMethods,
    author = {Ai, Wen-Yuan and Garbrecht, Bj\"orn and Tamarit, Carlos},
    title = "{Functional methods for false vacuum decay in real time}",
    eprint = "1905.04236",
    archivePrefix = "arXiv",
    primaryClass = "hep-th",
    doi = "10.1007/JHEP12(2019)095",
    journal = "JHEP",
    volume = "2019",
    number = {12},
    pages = "095",
    year = "2019"
}

@MISC{BlumRealTimeTunneling,
    author = "Blum, Kfir and Rosner, Omri",
    title = "{Unraveling the bounce: a real time perspective on tunneling}",
    eprint = "2309.07585",
    archivePrefix = "arXiv",
    primaryClass = "quant-ph",
    month = "9",
    year = "2023"
}

@article{SteingasserRealTimeInstantons,
    author = "Steingasser, Thomas and Kaiser, David I.",
    title = "{Toward quantum tunneling from excited states: Recovering imaginary-time instantons from a real-time analysis}",
    eprint = "2402.00099",
    archivePrefix = "arXiv",
    primaryClass = "hep-th",
    journal = {Phys. Rev. D},
    volume = {111},
    number = {9},
    pages = {096009},
    numpages = {17},
    month = "5",
    year = "2025",
    publisher = {American Physical Society},
    doi = {10.1103/PhysRevD.111.096009}
}

@article{LawrenceRealTimeTunneling,
    author = "Lawrence, Joseph E.",
    title = "{Semiclassical instanton theory for reaction rates at any temperature: How a rigorous real-time derivation solves the crossover temperature problem}",
    eprint = "2409.02820",
    archivePrefix = "arXiv",
    primaryClass = "physics.chem-ph",
    doi = "10.1063/5.0237368",
    journal = "J. Chem. Phys.",
    volume = "161",
    number = "18",
    pages = "184115",
    year = "2024"
}

@article{PatrascioiuComplexTime,
    author = "Patrascioiu, Adrian",
    title = "{Complex time and the Gaussian approximation}",
    reportNumber = "LA-UR-81-793",
    doi = "10.1103/PhysRevD.24.496",
    journal = "Phys. Rev. D",
    volume = "24",
    number = {2},
    pages = {496--504},
    year = "1981",
    month = {Jul},
    publisher = {American Physical Society}
}

@article{GamowAlphaDecay,
    author={Gamow, G.},
    title="{Zur Quantentheorie des Atomkernes}",
    journal={Zeitschrift f{\"u}r Physik},
    year={1928},
    month={Mar},
    day={01},
    volume={51},
    number={3},
    pages={204--212},
    issn={0044-3328},
    doi={10.1007/BF01343196}
}

@article{SiegertRadiativeStates,
    author = "Siegert, A. J. F.",
    title = "{On the Derivation of the Dispersion Formula for Nuclear Reactions}",
    doi = "10.1103/PhysRev.56.750",
    journal = "Phys. Rev.",
    volume = "56",
    number = {8},
    pages = "750--752",
    year = "1939",
    month = {Oct},
    publisher = {American Physical Society}
}

@article{IntroductionGamovVectors,
    author = "de la Madrid, R. and Gadella, M.",
    title = "{A pedestrian introduction to Gamow vectors}",
    eprint = "quant-ph/0201091",
    archivePrefix = "arXiv",
    doi = "10.1119/1.1466817",
    journal = "Am. J. Phys.",
    volume = "70",
    number = {6},
    pages = "626--638",
    year = "2002",
    month = {06}
}

@article{ResonancesIntroduction,
    author = "Hatano, Naomichi and Sasada, Keita and Nakamura, Hiroaki and Petrosky, Tomio",
    title = "{Some properties of the resonant state in quantum mechanics and its computation}",
    eprint = "0705.1388",
    archivePrefix = "arXiv",
    primaryClass = "quant-ph",
    doi = "10.1143/PTP.119.187",
    journal = "Prog. Theor. Phys.",
    volume = "119",
    number = {2},
    pages = "187--222",
    year = "2008",
    month = {02}
}

@article{BenderAnharmonicOscillator2,
    author = "Bender, Carl M. and Wu, T. T.",
    title = {{Anharmonic Oscillator. II. A Study of Perturbation Theory in Large Order}},
    doi = "10.1103/PhysRevD.7.1620",
    journal = "Phys. Rev. D",
    volume = "7",
    number = {6},
    pages = "1620--1636",
    year = "1973",
    month = {Mar},
    publisher = {American Physical Society}
}

@book{MarinoAdvancedQM,
    author = "Mari\~no, Marcos",
    title = "{Advanced Topics in Quantum Mechanics}",
    doi = "10.1017/9781108863384",
    isbn = "978-1-108-86338-4, 978-1-108-49587-5",
    publisher = "Cambridge University Press",
    year = "2021"
}

@article{FeldbruggeRealTimeTunneling,
    author = "Feldbrugge, Job and Jow, Dylan L. and Pen, Ue-Li",
    title = "{Complex classical paths in quantum reflections and tunneling}",
    eprint = "2309.12420",
    archivePrefix = "arXiv",
    primaryClass = "quant-ph",
    doi = "10.1103/PhysRevD.111.085027",
    journal = "Phys. Rev. D",
    volume = "111",
    number = "8",
    pages = "085027",
    year = "2025"
}

@article{SteingasserFiniteTemp,
    author = {Steingasser, Thomas and K\"onig, Morgane and Kaiser, David I.},
    title = "{Finite-temperature instantons from first principles}",
    eprint = "2310.19865",
    archivePrefix = "arXiv",
    primaryClass = "hep-th",
    doi = "10.1103/PhysRevD.110.L111902",
    journal = "Phys. Rev. D",
    volume = "110",
    number = "11",
    pages = "L111902",
    year = "2024"
}

@article{KhalfinLateTimeBehavior,
	author  = {Khalfin, L. A.},
	title   = "{Contribution to the Decay Theory of a Quasi-Stationary State}",
	url     = {http://www.jetp.ras.ru/cgi-bin/dn/e_006_06_1053.pdf},
	journal = {Soviet Phys. JETP},
	volume  = {6},
    pages = {1053--1063},
	year    = {1958},
	month   = {06}
}

@article{PeresDecayLaw,
    author = "Peres, Asher",
    title = "{Nonexponential decay law}",
    doi = "10.1016/0003-4916(80)90288-2",
    journal = "Annals Phys.",
    volume = "129",
    number = {1},
    pages = "33--46",
    year = "1980"
}

@article{BariBiorthogonalBasis,
    author = "N. K. Bari",
    title = "{Biorthogonal systems and bases in Hilbert space}",
    journal = "Uch. Zap. Mosk. Gos. Univ.",
    volume = "148",
    pages = "69--107",
    year = "1951",
	url = {https://www.mathnet.ru/eng/uzmu70}
}

@article{SternheimBiorthogonalBasis,
	title = {Non-Hermitian Hamiltonians, Decaying States, and Perturbation Theory},
	author = {Sternheim, Morton M. and Walker, James F.},
	journal = {Phys. Rev. C},
	volume = {6},
	number = {1},
	pages = {114--121},
	numpages = {0},
	year = {1972},
	month = {Jul},
	publisher = {American Physical Society},
	doi = {10.1103/PhysRevC.6.114},
	url = {https://link.aps.org/doi/10.1103/PhysRevC.6.114}
}

@article{BrodyBiorthogonalBasis,
	doi = {10.1088/1751-8113/47/3/035305},
	year = {2013},
	month = {dec},
	publisher = {IOP Publishing},
	volume = {47},
	number = {3},
	pages = {035305},
	author = {Brody, Dorje C.},
	title = {Biorthogonal quantum mechanics},
	journal = {J. Phys. A-Math.},
	eprint = "1308.2609",
    archivePrefix = "arXiv",
    primaryClass = "quant-ph",
}

@article{GrahamPathIntegralMethods1,
	title = "{Lagrangian for Diffusion in Curved Phase Space}",
	author = {Graham, Robert},
	journal = {Phys. Rev. Lett.},
	volume = {38},
	number = {2},
	pages = {51--53},
	numpages = {0},
	year = {1977},
	month = {Jan},
	publisher = {American Physical Society},
	doi = {10.1103/PhysRevLett.38.51},
	url = {https://link.aps.org/doi/10.1103/PhysRevLett.38.51}
}

@article{GrahamPathIntegralMethods2,
	author={Graham, Robert},
	title={Path integral formulation of general diffusion processes},
	journal={Z. Phys. B},
	year={1977},
	month={Sep},
	day={01},
	volume={26},
	number={3},
	pages={281--290},
	issn={1431-584X},
	doi={10.1007/BF01312935},
	url={https://doi.org/10.1007/BF01312935}
}

@article{dePireyPathIntegralMethods,
	author = {Thibaut Arnoulx de Pirey, Leticia F. Cugliandolo, Vivien Lecomte and Frédéric van Wijland},
	title = {Path integrals and stochastic calculus},
	journal = {Adv. Phys.},
	eprint = "2211.09470",
    archivePrefix = "arXiv",
    primaryClass = "cond-mat.stat-mech",
	volume = {71},
	number = {1--2},
	pages = {1--85},
	year = {2022},
	publisher = {Taylor \& Francis},
	doi = {10.1080/00018732.2023.2199229},
	URL = {https://doi.org/10.1080/00018732.2023.2199229},
}

@book{LangoucheFunctionalIntegration,
	title="{Functional Integration and Semiclassical Expansions}",
	author="{Langouche, F. and Roekaerts, D. and Tirapegui, E.}",
	year={1982},
	doi={10.1007/978-94-017-1634-5},
	series={Mathematics and Its Applications (MAIA)},
	volume={10},
	publisher={Springer Dordrecht}
}

@book{SibuyaEigenvalueProblems,
author = {Sibuya, Yasutaka},
isbn = {072042609X},
publisher = {North-Holland},
series = {North Holland Mathematics Studies 18},
title = {Global theory of a second order linear ordinary differential equation with a polynomial coefficient},
year = {1975}
}

@article{EremenkoODE,
year = {2011},
volume = {11},
number = {3},
number = {27},
pages = {473--503},
eprint = "1005.1696",
archivePrefix = "arXiv",
primaryClass = "math-ph",
author = {Alexandre Eremenko, Andrei Gabrielov},
title = {Singular perturbation of polynomial potentials with applications to $\mathcal{PT}$-symmetric families},
journal = {Mosc. Math. J.}
}

@article{ShinPTEigenvalues,
    author = {Shin, K. C.},
    title = {On the eigenproblems of $\mathcal{PT}$-symmetric oscillators},
    journal = {J. Math. Phys.},
    volume = {42},
    number = {6},
    pages = {2513--2530},
    year = {2001},
    month = {06},
    issn = {0022-2488},
    doi = {10.1063/1.1366328},
    url = {https://doi.org/10.1063/1.1366328},
    eprint = {math-ph/0007006},
    archivePrefix = "arXiv"
}

@article{BenderAnharmonicOscillator,
  title = {Anharmonic Oscillator},
  author = {Bender, Carl M. and Wu, Tai Tsun},
  journal = {Phys. Rev.},
  volume = {184},
  number = {5},
  pages = {1231--1260},
  numpages = {0},
  year = {1969},
  month = {Aug},
  publisher = {American Physical Society},
  doi = {10.1103/PhysRev.184.1231},
  url = {https://link.aps.org/doi/10.1103/PhysRev.184.1231}
}

@article{BenderAnalyticylContinuation,
    author = "Bender, Carl M. and Turbiner, Alexander",
    title = "{Analytic continuation of eigenvalue problems}",
    reportNumber = "WU-HEP-92-13",
    doi = "10.1016/0375-9601(93)90153-Q",
    journal = "Phys. Lett. A",
    volume = "173",
    number = {6},
    pages = "442--446",
    year = "1993"
}

@article{MostafazadehComplexContour,
    author = "Mostafazadeh, Ali",
    title = "{Pseudo-Hermitian description of $\mathcal{PT}$-symmetric systems defined on a complex contour}",
    eprint = "quant-ph/0410012",
    archivePrefix = "arXiv",
    doi = "10.1088/0305-4470/38/14/011",
    journal = "J. Phys. A",
    volume = "38",
    number = {14},
    pages = "3213--3234",
    year = "2005"
}

@article{MostafazadehReview,
    author = "Mostafazadeh, Ali",
    title = "{Pseudo-Hermitian Representation of Quantum Mechanics}",
    eprint = "0810.5643",
    archivePrefix = "arXiv",
    primaryClass = "quant-ph",
    doi = "10.1142/S0219887810004816",
    journal = "Int. J. Geom. Meth. Mod. Phys.",
    volume = "7",
    pages = "1191--1306",
    year = "2010"
}

@book{FedoryukAsymptoticAnalysis,
  title={Asymptotic Analysis: Linear Ordinary Differential Equations},
  author={Mikhail V. Fedoryuk},
  year={1993},
  publisher={Springer Berlin}
}

@article{WagnerExcitedStateTunneling,
    author = {Garbrecht, Bj\"orn and Wagner, Nils},
    title = "{False vacuum decay of excited states in finite-time instanton calculus}",
    eprint = "2412.20431",
    archivePrefix = "arXiv",
    primaryClass = "hep-th",
    doi = "10.1007/JHEP05(2025)076",
    journal = "JHEP",
    volume = "2025",
    number={5},
    pages = "076",
    year = "2025"
}

@article{HertzbergRealTimeTunneling,
    author = "Hertzberg, Mark P. and Yamada, Masaki",
    title = "{Vacuum Decay in Real Time and Imaginary Time Formalisms}",
    eprint = "1904.08565",
    archivePrefix = "arXiv",
    primaryClass = "hep-th",
    doi = "10.1103/PhysRevD.100.016011",
    journal = "Phys. Rev. D",
    volume = "100",
    number = "1",
    pages = "016011",
    year = "2019"
}

@article{MatsuiRealTimePI,
    author = "Matsui, Hiroki",
    title = "{Lorentzian path integral for quantum tunneling and WKB approximation for wave-function}",
    eprint = "2102.09767",
    archivePrefix = "arXiv",
    primaryClass = "gr-qc",
    doi = "10.1140/epjc/s10052-022-10374-1",
    journal = "Eur. Phys. J. C",
    volume = "82",
    number = "5",
    pages = "426",
    year = "2022"
}

@article{MouRealTimeTunneling,
    author = "Mou, Zong-Gang and Saffin, Paul M. and Tranberg, Anders",
    title = "{Quantum tunnelling, real-time dynamics and Picard-Lefschetz thimbles}",
    eprint = "1909.02488",
    archivePrefix = "arXiv",
    primaryClass = "hep-th",
    doi = "10.1007/JHEP11(2019)135",
    journal = "JHEP",
    volume = "2019",
    number = "11",
    pages = "135",
    year = "2019"
}

@article{SherVacuumStabilitySM,
    author = "Sher, Marc",
    title = "{Electroweak Higgs Potentials and Vacuum Stability}",
    reportNumber = "WU-TH-88-8",
    doi = "10.1016/0370-1573(89)90061-6",
    journal = "Phys. Rept.",
    volume = "179",
    number = {5},
    pages = "273--418",
    year = "1989"
}

@article{Espinosa_VacSStability,
    author = "Espinosa, J. R. and Quirós, M.",
    title = "{Improved metastability bounds on the standard model Higgs mass}",
    eprint = "hep-ph/9504241",
    archivePrefix = "arXiv",
    reportNumber = "CERN-TH-95-18, CERN-TH-95-018, DESY-95-039, IEM-FT-97-95",
    doi = "10.1016/0370-2693(95)00572-3",
    journal = "Phys. Lett. B",
    volume = "353",
    number = {2},
    pages = "257--266",
    year = "1995"
}

@article{FondaDecayTheory,
    author = "Fonda, L. and Ghirardi, G. C. and Rimini, A.",
    title = "{Decay theory of unstable quantum systems}",
    doi = "10.1088/0034-4885/41/4/003",
    journal = "Rept. Prog. Phys.",
    volume = "41",
    number = {4},
    pages = "587--631",
    year = "1978",
    month = {apr}
}

@MISC{Wagner_CC_Resolution,
    author = {Garbrecht, Bj{\"o}rn and Wagner, Nils},
    title = "{Path integral analysis of Schr{\"o}dinger-type eigenvalue problems in the complex plane: Establishing the relation between instantons and resonant states}",
    eprint = "2507.23125",
    archivePrefix = "arXiv",
    primaryClass = "hep-th",
    month = "7",
    year = "2025"
}

@article{Ai_PT_PathInt,
    author = "Ai, Wen-Yuan and Bender, Carl M. and Sarkar, Sarben",
    title = "{$\mathcal{PT}$-symmetric $-g\varphi^4$ theory}",
    eprint = "2209.07897",
    archivePrefix = "arXiv",
    primaryClass = "hep-th",
    reportNumber = "KCL-PH-TH-2022-25",
    doi = "10.1103/PhysRevD.106.125016",
    journal = "Phys. Rev. D",
    volume = "106",
    number = "12",
    pages = "125016",
    year = "2022"
}

@article{Bender_PT_PathInt,
    author = "Bender, Carl M. and Brody, Dorje C. and Chen, Jun-Hua and Jones, Hugh F. and Milton, Kimball A. and Ogilvie, Michael C.",
    title = "{Equivalence of a Complex $\mathcal{PT}$-Symmetric Quartic Hamiltonian and a Hermitian Quartic Hamiltonian with an Anomaly}",
    eprint = "hep-th/0605066",
    archivePrefix = "arXiv",
    doi = "10.1103/PhysRevD.74.025016",
    journal = "Phys. Rev. D",
    volume = "74",
    pages = "025016",
    year = "2006"
}

@MISC{SteingasserExcitedStateSteadyonslong,
    author = "Lin, Joshua and Scheihing-Hitschfeld, Bruno and Steingasser, Thomas",
    title = "{Quantum tunneling from excited states in the steadyon picture}",
    eprint = "2511.08670",
    archivePrefix = "arXiv",
    primaryClass = "hep-th",
    month = "11",
    year = "2025"
}

@MISC{SteingasserExcitedStateSteadyonsshort,
    author = "Lin, Joshua and Scheihing-Hitschfeld, Bruno and Steingasser, Thomas",
    title = "{Path integral predictions for pre-asymptotic false vacuum decay}",
    eprint = "2511.08669",
    archivePrefix = "arXiv",
    primaryClass = "hep-th",
    month = "11",
    year = "2025"
}

\end{document}